\documentclass[aps,prd,twocolumn,showpacs,eqsecnum,floatfix]{revtex4-1}

\usepackage{graphicx}
\usepackage{amsmath}
\usepackage{amssymb}
\usepackage{hyperref}

\def\apj{Astrophys.\ J.\ }

\def\aap{Astron.\ \& Astrophys.\ }

\def\mnras{Mon.\ Not.\ Roy.\ Astron.\ Soc.\ }

\font\FermiSmallfont=cmssq8 scaled 1200

\def\UMDppthead#1#2#3{
\null 
\begin{center}\vskip -1.0truein{\hbox to 7.5truein {
\hfill
\vbox to 1in {\vfill \FermiSmallfont
              \hbox{#1}
              \hbox{#2}
              \hbox{#3}
              \vfill}
}}\vskip-0.0truein\end{center}}

\begin{document}

\UMDppthead{UMD-PP-10-020}{arXiv:1011.5090}{}

\title{Current and Future Constraints on Dark Matter from Prompt and
  Inverse-Compton Photon Emission in the Isotropic Diffuse Gamma-ray
  Background}

\author{Kevork N.\ Abazajian$^{1,2}$} \email{kevork@uci.edu}
\author{Steve Blanchet$^{2,3}$} \email{steve.blanchet@epfl.ch}
\author{J.\ Patrick Harding$^{1,2}$} \email{hard0923@umd.edu}
\affiliation{$^1$Department of Physics and Astronomy, University of
  California, Irvine, Irvine, California 92697 USA}
\affiliation{$^2$Maryland Center for Fundamental Physics \& Joint
  Space-Science Institute, Department of Physics, University of
  Maryland, College Park, Maryland 20742 USA}
\affiliation{$^3$Instituto de F\'isica Te\'orica, IFT-UAM/CSIC Nicolas
  Cabrera 15, UAM Cantoblanco, 28049 Madrid, Spain}

\date{\today}

\begin{abstract}
  We perform a detailed examination of current constraints on
  annihilating and decaying dark matter models from both prompt and
  inverse-Compton emission photons, including both model-dependent and
  model-independent bounds.  We also show that the observed isotropic
  diffuse gamma-ray background (DGRB), which provides one of the most
  conservative constraints on models of annihilating weak-scale dark
  matter particles, may enhance its sensitivity by a factor of $\sim$
  2 to 3 (95\% C.L.) as the Fermi-LAT experiment resolves DGRB
  contributing blazar sources with five years of observation. For our
  forecasts, we employ the results of constraints to the
  luminosity-dependent density evolution plus blazar spectral energy
  distribution sequence model, which is constrained by the DGRB and
  blazar source count distribution function.
\end{abstract}

 
\maketitle

\section{Introduction}
The existence of cosmological dark matter is well established by
observations of galaxy clusters, galaxy rotation curves, the cosmic
microwave background (CMB) and large-scale structure, though its
nature remains a fundamental problem in cosmology and particle
physics.  There exists an abundance of particle candidates which could
account for the dark matter (for a review, see,
e.g. \cite{Feng:2010gw}).  For a class of particles with weak-scale
interaction and weak-scale particle mass, weakly interacting massive
particles (WIMPs), their production in the early Universe in thermal
processes naturally produces the observed dark matter density, largely
independent of the particle mass.  

Thermal freeze-out predicts a canonical annihilation rate of $\langle
\sigma_{\rm A} v\rangle \approx 3\times 10^{-26}\rm\ cm^3\ s^{-1}$.
This predicted annihilation rate in standard model channels leads to
energetic gamma-ray production in the hadronization of quarks, the
Higgs or gauge bosons, through bremsstrahlung in the case of the
lighter leptons, or directly to two gammas through higher order
processes.  The diffuse gamma-ray background (DGRB) was forecast to be
one of the most robust constraints on annihilating WIMP dark matter
\cite{Baltz:2008wd}.  Because of a more conservative model for the
extragalactic dark matter signal, the conservative limits on dark
matter annihilation presented by the Fermi-LAT Collaboration from the
1-year observation of the DGRB \cite{Abdo:2010nz,Abdo:2010dk} were
weaker than prelaunch estimates \cite{Baltz:2008wd}.  On the other
hand, it was shown that the DGRB has an irreducible contribution from
the Milky Way Galactic dark matter halo \cite{Mack:2008wu} that is
greater in amplitude than the conservative estimates of the
extragalactic contribution, and correspondingly has more stringent
limits over many annihilation channels \cite{Abazajian:2010sq}.  This
irreducible, isotropic component is due to the fact that annihilating
or decaying dark matter in the Milky Way halo has an isotropic
component equal to the minimum of the annihilation or decay signal.
This minimum is equal to the amount toward the anti-Galactic Center.

The isotropic DGRB has several potential astrophysical source
contributions, including
blazars~\cite{Stecker:1996ma,Inoue:2008pk,Singal:2011yi,Cavadini:2011ig},
starburst galaxies \cite{Fields:2010bw} and millisecond pulsars
\cite{FaucherGiguere:2009df}.  The only model that successfully
predicts the shape and amplitude of the DGRB over all energies is the
luminosity-dependent density evolution (LDDE) blazar spectral energy
density (SED) sequence model with an active galactic nuclei (AGN)
contribution \cite{Inoue:2010tj,Abazajian:2010pc}.  The SED-sequence
model matches the shape of the observed blazar SED's luminosity
dependence \cite{Fossati:1997vu,*Fossati:1998zn,*Donato:2001ge}.
Reference~\cite{Inoue:2010tj} reproduces well the DGRB as observed by
Fermi-LAT, while several other analyses under-produce the DGRB from
blazars.  Reference~\cite{Inoue:2010tj} estimates that $\gtrsim 98\%$ of
the blazar flux contributing to the DGRB will be resolved in the
5-year Fermi-LAT survey.  Prior work that under-produces the DGRB uses
a single power-law for the spectrum of all blazars instead of the
observed SED sequence for blazars, e.g~\cite{Collaboration:2010gqa}.
Other recent work with varied blazar population models, including
spectral shape variation~\cite{Venters:2011gg}, possible point source
confusion~\cite{Stecker:2010di}, and BL Lac dominance of the
unresolved portion~\cite{Neronov:2011kg} also find that a substantial
portion of the DGRB could arise from the blazar population.

There may also be an unmodeled, unreduced isotropic Galactic component
to the DGRB~\cite{Abdo:2010nz,Abdo:2010dk}.  It should be noted that
two things could happen if there is a presently unremoved Galactic
isotropic diffuse component: one, it is detected, modeled, and
removed, which will make future constraints stronger; or, two, it
remains a systematic diffuse background, which means our starting
assumption of an LDDE SED-sequence model is not the correct model for
the DGRB.  However, this uncertainty cannot be removed without further
observational analysis.

Below, we calculate and show the current constraints from the DGRB on
dark matter annihilation and decay gamma-rays from the prompt as well
as inverse-Compton components.  In addition, when adopting the LDDE
plus SED-sequence model forecasts of the Fermi-LAT resolved DGRB,
future observations will extend the reach of Fermi-LAT sensitivity to
dark matter typically by a factor of 2 to 3.  We explore in
detail the forecasts on standard WIMP dark matter and leptonic-channel
motivated models~\cite{ArkaniHamed:2008qn,*Pospelov:2008jd} including
Asymmetric Dark Matter models~\cite{Falkowski:2011xh,*Cai:2009ia},
nonthermal winolike dark matter ~\cite{Grajek:2008jb,*Kane:2009if},
and decaying dark matter \cite{Cirelli:2009dv}, as well as models of
light ($\sim$8 GeV) WIMP dark matter in scalar dark matter
models~\cite{Burgess:2000yq,Andreas:2008xy,*Arina:2010rb}.  The
resolution of the DGRB into point source blazars will reduce the DGRB
amplitude and ultimately enhance the limits on annihilation in WIMP
dark matter models.

In a companion paper, Ref.~\cite{Abazajian:2010pc} (ABH2), we
constrain our adopted SED-sequence model using the observed DGRB
spectrum as well as the observed blazar source count distribution
function, $dN/dF$. In agreement with Ref.~\cite{Inoue:2010tj}, ABH2
found that $\gtrsim 95\%$ of the flux from blazars will be resolved
with 5 years of Fermi-LAT observation.  The resolution of the DGRB is
in fact similar to the resolution of the cosmic x-ray background
observed by {\it Chandra}, which in turn provided stringent
constraints on decaying light sterile neutrino dark matter
\cite{Abazajian:2006jc}.  

The work presented here advances previous work on prompt gamma-ray
emission in annihilating dark matter (e.g., \cite{Abazajian:2010sq})
by including the enhanced constraints and sensitivity from
inverse-Compton (IC) emission present in the Fermi-LAT observation of
the DGRB as well as forecasts of the improvement of this
sensitivity. In addition, we go beyond previous analyses of
IC emission enhancement of the extragalactic and Galactic
signals (e.g., \cite{Cirelli:2009dv,Profumo:2009uf}) by applying the
IC enhancement in the Fermi-LAT observation of the DGRB and its
forecast improvement.

\section{The Blazar Population and SED-sequence model}

The SED-sequence model specifies the cosmological blazar spatial
distribution and spectrum for a given blazar luminosity.  It is based
on the observed evolution of the peak flux in synchrotron and IC
emission with luminosity.  The luminosity-dependent density evolution
model specifies the gamma-ray luminosity function of blazars through a
fraction of the total AGN population and its x-ray luminosity function.
Our blazar population and SED-sequence model in ABH2 successfully
reproduces the observed DGRB and blazar source count $dN/dF$.

Our model is a modification of that by Inoue and
Totani~\cite{Inoue:2008pk}, and is detailed in ABH2. We provide a
summary here.  The bolometric blazar jet luminosity $P$ and disk x-ray
luminosity $L_X$ are related by $P=10^q L_X$.  The blazar gamma-ray
redshift-dependent luminosity function is given as a fraction $\kappa$
of the AGN x-ray luminosity function (XLF),
$\rho_\gamma(L_\gamma,z)=\kappa (dL_X/dL_\gamma) \rho_X(L_X,z).$ We
adopt the AGN XLF of Ueda et al.~\cite{Ueda:2003}.  The main fit
parameter in the XLF is the faint-end slope index, $\gamma_1$.  The
model also includes a nonblazar AGN component which dominates at
lower energies, $E_\gamma \lesssim 1\rm\ GeV$. 

In ABH2, we constrain the blazar population model by simultaneously
fitting the DGRB spectrum as well as the blazar flux source count
distribution function $dN/dF$ observed by Fermi-LAT
\cite{Abdo:2010nz,Collaboration:2010gqa}.  The best fit parameters we
find are $q = 4.19^{+0.57}_{-0.13}$, $\log(\kappa/10^{-6}) =
0.38^{+0.15}_{-0.70}$, and $\gamma_1 = 1.51^{+0.10}_{-0.09}$.  These
are consistent with previous work \cite{Inoue:2008pk}, though more
constrained because we are also fitting the source count distribution
function $dN/dF$.  The model reproduces the DGRB and blazar $dN/dF$,
with a reduced $\chi^2/{\rm DOF} = 0.63$.

Using the $dN/dF$ estimated from a power-law blazar spectrum model is
not perfect, since the efficiency depends on this model
\cite{Collaboration:2010gqa}.  However,
Ref.~\cite{Collaboration:2010gqa} tested the $dN/dF$ estimate with a
non-power-law fit to the blazar spectra and found it did not
appreciably change the estimates of $dN/dF$, adding a systematic
uncertainty of $10\%$.  We also checked this sensitivity with
a test fit by increasing the errors on the measured $dN/dF$ at low
flux and our model did not prefer a different amplitude or shape to
the source counts at the low flux where the efficiency for blazar
detection is low.

\begin{figure}[t]
\includegraphics[width=3.4truein]{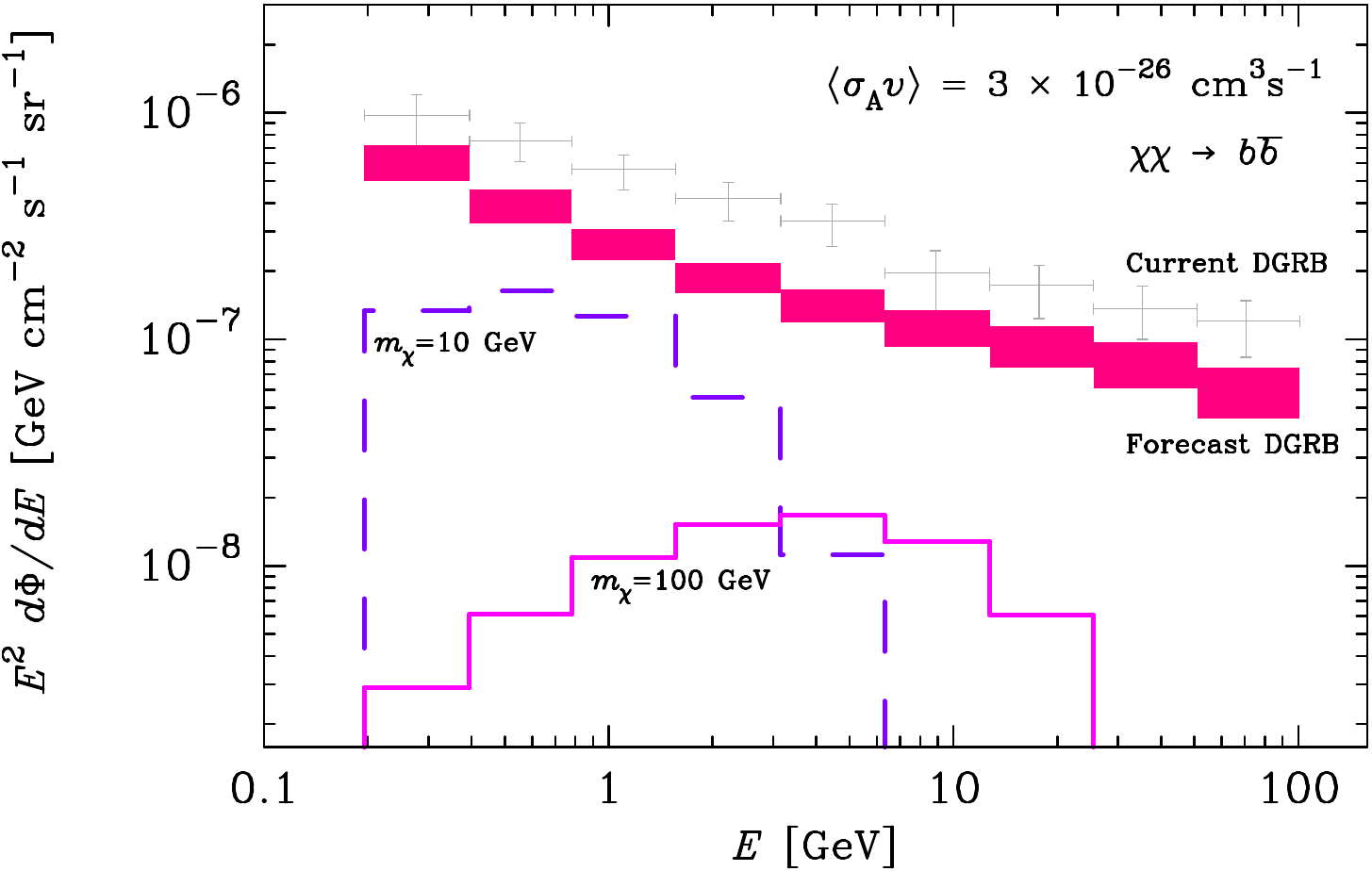}
\vskip -0.2 cm
\caption{\small Shown are the current Fermi-LAT observed DGRB
  \cite{Abdo:2010nz} in grey, and the forecast DGRB upper and lower 
  95\% C.L.\ central values as the boxed (magenta) regions.   Also shown are the 
  expected emission from the Galactic and extragalactic contribution for WIMP
  annihilation into $b \bar b$ for the WIMP particle masses 10 GeV and
  100 GeV, for the canonical 
  $\langle \sigma_{\rm A} v\rangle = 3\times 10^{-26}\rm\ cm^3\ s^{-1}$.
  \label{spectrum}}
\end{figure}

Our model fits the current DGRB, and, furthermore, predicts the DGRB
for the expected enhanced sensitivity to point sources after 5 years
of Fermi-LAT data, $2 \times 10^{-9}\rm\ photons\ cm^{-2}\ s^{-1}$,
which will resolve $94.7^{+1.9}_{-2.1}\%$ of the flux from blazars.
This expected enhanced point source sensitivity value is the Fermi-LAT
Collaboration's estimate of the LAT flux sensitivity to point sources
at high-latitude with gamma-ray index of $\sim$2
\cite{Atwood:2009ez}.\footnote{http://fermi.gsfc.nasa.gov/science/resources/aosrd/} In ABH2, we
find the 68\% and 95\% confidence level (C.L.) upper and lower limit
forecasts for the DGRB $E^2 d\Phi/dE$ when varying the fit parameters.
This model finds that the DGRB will reduce by a factor of 1.6-2.6
(95\% C.L.) with the spatial point-source resolution of the blazar
contribution, after five years of the Fermi-LAT mission. The current
and forecast spectra are shown in Fig.~\ref{spectrum}.  In our 
forecasts, we calculate the limits using the flux as predicted by the
model at the minimal and maximal values, not simply performing a
scaling of the limits.

From the first to second Fermi source catalogs, there were 162
potentially spurious sources designated, indicating the sources'
further identification with a spatially extended source, source
variability, or other systematic effects \cite{Abdo:2011bm}.  The
catalogs include sources with the test statistic $TS=25$, while the
Fermi-LAT DGRB analysis only removed sources with $TS=50$, or with
higher significance.  (For the definition of $TS$, see Eq.~20 of
Ref.~\cite{Mattox:1996zz}.)  This type of spurious contamination may alter
forecasts for the DGRB, though the higher significance required for
the exclusion of sources in the DGRB spectrum would likely reduce or
eliminate this systematic effect.

\section{Current and Future Sensitivity to Dark Matter Models of the
  Fermi-LAT DGRB}

The signal constrained by observations of the DGRB is the annihilation
or decay of dark matter both in the Milky Way Galaxy and
extragalactically.  There is an irreducible contribution to the
background from Galactic annihilation or decay that is isotropic and
equal, at minimum, to the emission from the Anti-Galactic-Center
(AGC).  Here we examine in detail constraints on annihilating and
decaying dark matter from both prompt and IC emission of photons.

\subsection{Diffuse Emission from Annihilating Dark Matter}

The products of dark matter annihilation emit in gamma-rays in several
ways: annihilation channels that include the direct emission of a
photon, decay of annihilation products into photons, and IC scattering
of daughter electrons off of background radiation. Through loop
contributions to the annihilation cross section of dark matter, it is
possible to have a direct $\gamma\gamma$ line signal, through a
typically small branching fraction. If the dark matter annihilation
particles include hadrons, then the decay chain of the products will
lead to neutral pion decay into photons. Additionally, electrons among
the dark matter annihilation daughters will up-scatter background
photons from the cosmic microwave background (CMB) and starlight to
gamma-ray energies through IC scattering. We will consider the latter
two cases in this paper: photons from prompt radiation in
bremsstrahlung and hadronization, and IC emission from electron
daughter particles. We calculate the photon and electron spectrum from
annihilation using the software {\sc PYTHIA}~\cite{Sjostrand:2006za}. To be
conservative, we only consider IC emission from the CMB, not starlight
or the infrared (IR) background, which is a good approximation in the
direction of the AGC that we will consider.

There are two sources of dark matter that contribute to the DGRB,
Galactic and extragalactic dark matter. The signal from Galactic dark
matter is largest in the line-of-sight toward the Galactic Center and
is much smaller in other directions. However, there is an irreducible,
isotropic signal from the Galactic dark matter that is equal to the
signal from the AGC. This isotropic signal can be the
strongest dark matter contributor to the DGRB given conservative
assumptions about the extragalactic contribution~\cite{Mack:2008wu}.

To calculate the dark matter annihilation flux for a given
cross section $\langle\sigma_A v\rangle$ and photon spectrum
$dN_\gamma/dE$, we follow the treatment of
Ref.~\cite{Abazajian:2010sq}. The
Galactic contribution to the diffuse flux is given by:
\begin{eqnarray}
  \frac{d\Phi_\gamma}{dEd\Omega}&=&\frac{\langle\sigma_A v\rangle}{2}\frac{\mathcal{J}({\rm AGC})}{\rm J_0}\frac{1}{4\pi m_\chi^2} \frac{dN_\gamma}{dE},\label{DMannfluxeqn}\\
  \mathcal{J}({\rm
    AGC})&=&\frac{1}{\Delta\Omega_{\rm obs}}\int_{\Delta\Omega_{\rm obs}}{\mathcal{J}(0,
    180^\circ) d\Omega}\label{JAGCeqn},\\
  \mathcal{J}(b,\ell)&=&{\rm J_0} \int_{x_{\rm min}}^{x_{\rm
      max}}{\rho^2 \left(r_{\rm gal}(b,\ell,x)\right)dx},\label{rhosqinteqn}\\
  r_{\rm gal}(b,\ell,x)&=&\sqrt{R_\odot^2-2x R_\odot\cos(\ell)\cos(b)+x^2},\label{rbleqn}
\end{eqnarray}
evaluated conservatively at the AGC ($b=0^\circ$,
$\ell=180^\circ$). In these equations, $m_\chi$ is the dark matter
particle mass, $x$ is the line-of-sight distance, $R_\odot$ is the
distance from the Galactic Center to the sun, and ${\rm J_0} \equiv
1/[8.5\ \rm kpc (0.3\ GeV\ cm^{-3})^2]$ is an arbitrary constant that
cancels in the final expression for flux. For the Fermi-LAT, the solid
angle above $\lvert b\rvert>10^\circ$ has $\Delta\Omega_{\rm
  obs}=10.4$. The quantity $\mathcal{J}$ is the dark matter density
squared integrated along the line of sight, and $\mathcal{J}({\rm
  AGC})$ is this value averaged over the observed angular sky
region. For dark matter density $\rho$ we use the minimal Einasto
profile for the Milky Way halo:
\begin{equation}
\rho_{\rm Einasto}(r)=\rho_s
\exp\left[-\frac{2}{\alpha_E}\left(\left(\frac{r}{r_s}\right)^{\alpha_E}
    -1\right)\right],
\label{rho_einasto}
\end{equation}
with $\alpha_E=0.22$, $r_s = 21\rm\ kpc$, $r_\odot = 8.28\rm\ kpc$,
and $\rho_\odot = 0.385\rm\ GeV\ cm^{-3}$ as in
Ref.~\cite{Abazajian:2010sq}. This profile is a conservative choice
and gives $\mathcal{J}({\rm AGC})=0.62$, and extreme assumptions about
the Milky Way dark matter profile only change this value by $\sim
10\%$.

The Milky Way dark matter halo has abundant substructure which
enhances the annihilation rate of dark matter.  Following
Ref.~\cite{Kamionkowski:2010mi}, the boost factor for the annihilation
due to substructure is
\begin{eqnarray}
B(r) &=& f_s e^{\Delta^2}\nonumber \\
&\ & +(1-f_s) \frac{1+\alpha}{1-\alpha}\left[\left(\frac{\rho_{\rm
        max}}{\rho_h}\right)^{1-\alpha} -1\right].
\label{boostatr}
\end{eqnarray}
The fraction $f_s$ of the halo volume is filled with a smooth dark
matter component with density $\rho_h$. The maximal density of the PDF
$\rho_{\rm max}$ is taken to be the scale density $\rho_s$ of the
earliest forming halos.  The first term in Eq.~\eqref{boostatr}, $f_s
e^{\Delta^2}$, is due to the variation in the smooth component, which
contributes only a few percent to the boost, and therefore we ignore
it.  The second term is the boost factor due to substructure.  The
total luminosity boost due to the entire Galactic halo within radius
$R$ is
\begin{equation}
B(<R) = \frac{\int_0^R B(r) \rho(r)^2 r^2 dr}{\int_0^R \rho(r)^2 r^2
  dr}, \label{bofr}
\end{equation}
where $r$ is the halo-centric radial coordinate.  The annihilation
rate is larger in all directions than the AGC,
therefore we calculate the luminosity with boost along that
line of sight as the minimal annihilation rate due to our Galactic
halo.  Along the line-of-sight,
\begin{equation}
  \mathcal{J}_{\rm boost}(b,\ell)={\rm J_0} \int_{x_{\rm min}}^{x_{\rm
      max}}{B(r_{\rm gal}(b,\ell,x)) \rho^2 \left(r_{\rm
        gal}(b,\ell,x)\right)dx}.\label{boostJ}
\end{equation}
There is a partial reduction of the total luminosity boost
[Eq.~\eqref{bofr}] along the line of sight to the AGC due to our
presence within the Galactic halo.  Using the central value of $\alpha
= 0$ from simulations, the boost is $B_{\rm AGC} \equiv
\mathcal{J}_{\rm boost}({\rm AGC})/\mathcal {J}({\rm AGC}) = 3.3$.
Though the boost factor toward the Galactic Center is expected to be
unity~\cite{Kamionkowski:2010mi}, that from the total Galactic halo
can approach $\sim$20 to 2000.  Therefore, the approximation of a
total Galactic boost in the DGRB field of view as 3.3 is conservative.

In addition to this Galactic contribution to the dark matter flux, we
include a subdominant contribution from extragalactic dark matter
annihilations~\cite{Bergstrom:2001jj,Ullio:2002pj}. This contribution is given
by
\begin{eqnarray}
  \frac{d\Phi_\gamma}{dEd\Omega} &=&\frac{\langle\sigma_A v\rangle}{2}\ 
  \frac{c}{4\pi H_0}\ \frac{(f_{\rm DM} \Omega_m)^2\rho_{\rm
      crit}^2}{m_\chi^2} \times \label{EGdecay}\cr 
  && \!\! \int_0^{z_{\rm up}}\!{\frac{f(z) (1+z)^3 e^{-\tau(z,E^\prime)}}{\sqrt{(1+z)^3 \Omega_m + \Omega_\Lambda}}
    \frac{dN(E^\prime)}{dE^\prime}dz} ,
\end{eqnarray}
where $H_0$ is the Hubble constant, $\Omega_m$ is the matter density
in units of the critical density, $\rho_{\rm crit}$, and the fraction
of matter in dark matter is $f_{\rm DM} = \Omega_{\rm DM}/(\Omega_{\rm
  DM}+\Omega_b) \approx 0.833$, where we take the fraction of critical
density of the dark matter as $\Omega_{\rm DM} = 0.237$, and baryon
density $\Omega_b = 0.0456$~\cite{Komatsu:2010fb}. Here, $E^\prime=E(1+z)$
is the source energy of the photons, and $z_{\rm up}=(m_\chi/E)-1$ is
the maximum redshift to get a photon with energy $E$. The factor
$f(z)$ accounts for the increase in density squared during halo growth
and the redshift evolution of the halo mass function.  We adopt the
fit of Refs.~\cite{Ullio:2002pj,Ando:2005hr,*Yuksel:2007ac}, namely:
\begin{equation}
f(z) = f_0 10^{0.9\left[\exp\left(-0.9 z\right) -1\right] - 0.16 z}.
\label{fando}
\end{equation}
For the Einasto profile, $f_0 \simeq 3\times 10^4$. We also include
the boost factor of 6.6, from the total luminosity of a halo
$B(<R)$. This extragalactic contribution only accounts for $\lesssim
30\%$ of the total diffuse flux from dark matter.

To calculate the limits on $\langle\sigma v\rangle$, we attribute all
the DGRB to dark matter annihilation (or decay), including both the
Galactic and extragalactic dark matter contributions. This provides an
upper limit on the cross section of annihilating dark matter (or a
lower limit on the lifetime of decaying dark matter).  We use the
upper and lower 68\% and 95\% C.L.\ forecast fluxes corresponding to the
extremal upper and lower fluxes in the three-dimensional contour in
$q$, $\kappa$ and $\gamma_1$ parameter space, as constrained by the
DGRB spectrum and source count distribution function, as described in
our companion paper~\cite{Abazajian:2010pc}.  We take errors on the
forecast DGRB for all of the upper and lower 68\% and 95\% C.L.\ forecast
fluxes each to be proportional to its amplitude, which corresponds to
the modeling methods of the DGRB measurement, though this is not
necessarily the ultimate scaling of the errors.  To do so, a Monte
Carlo of the modeling methods of Ref.~\cite{Abdo:2010nz} would need to
be performed, which is beyond the scope of this work.

\begin{figure*}[t]
\includegraphics[width=3.4truein]{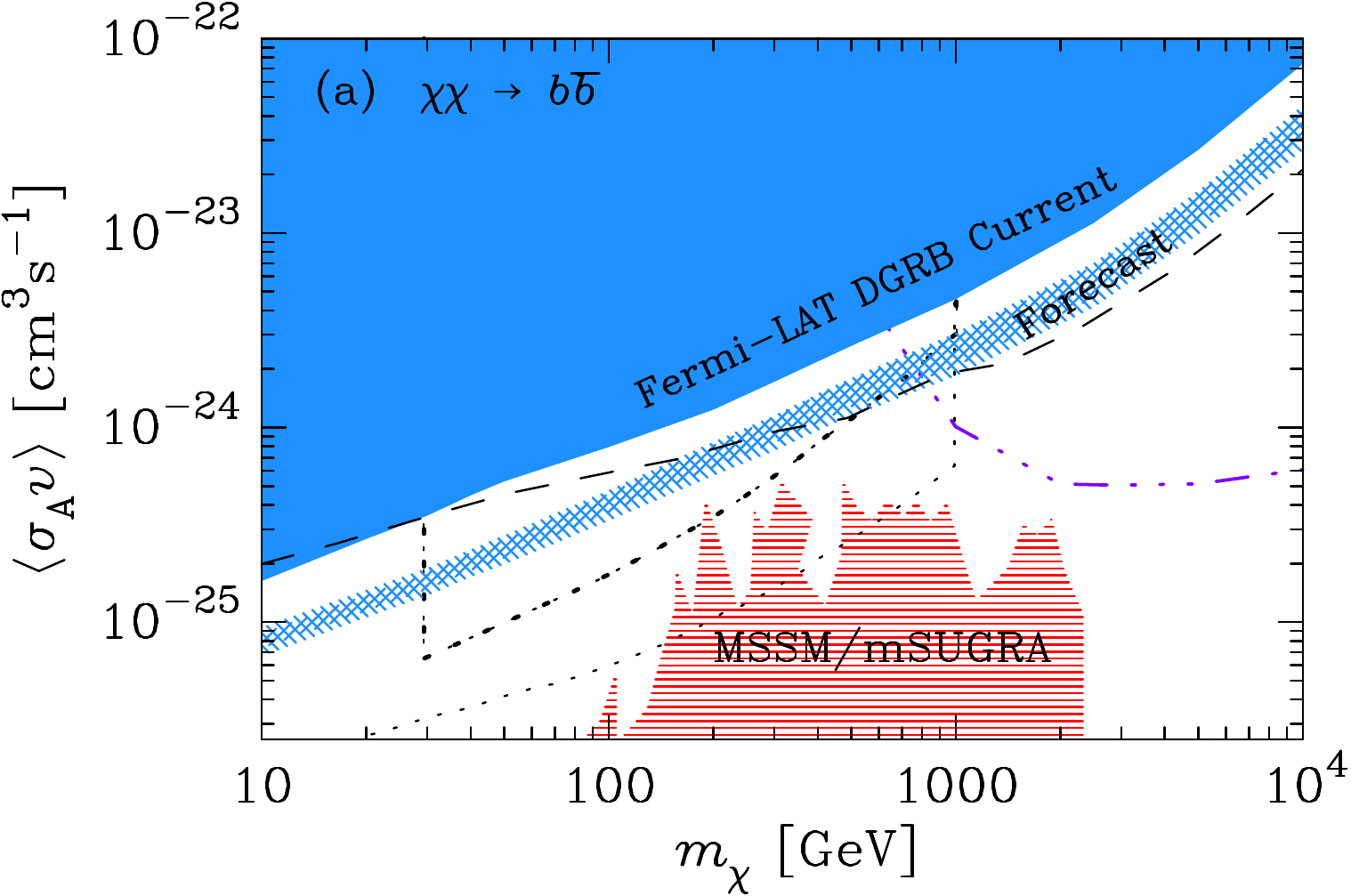} 
\hspace{0.3cm}
\includegraphics[width=3.4truein]{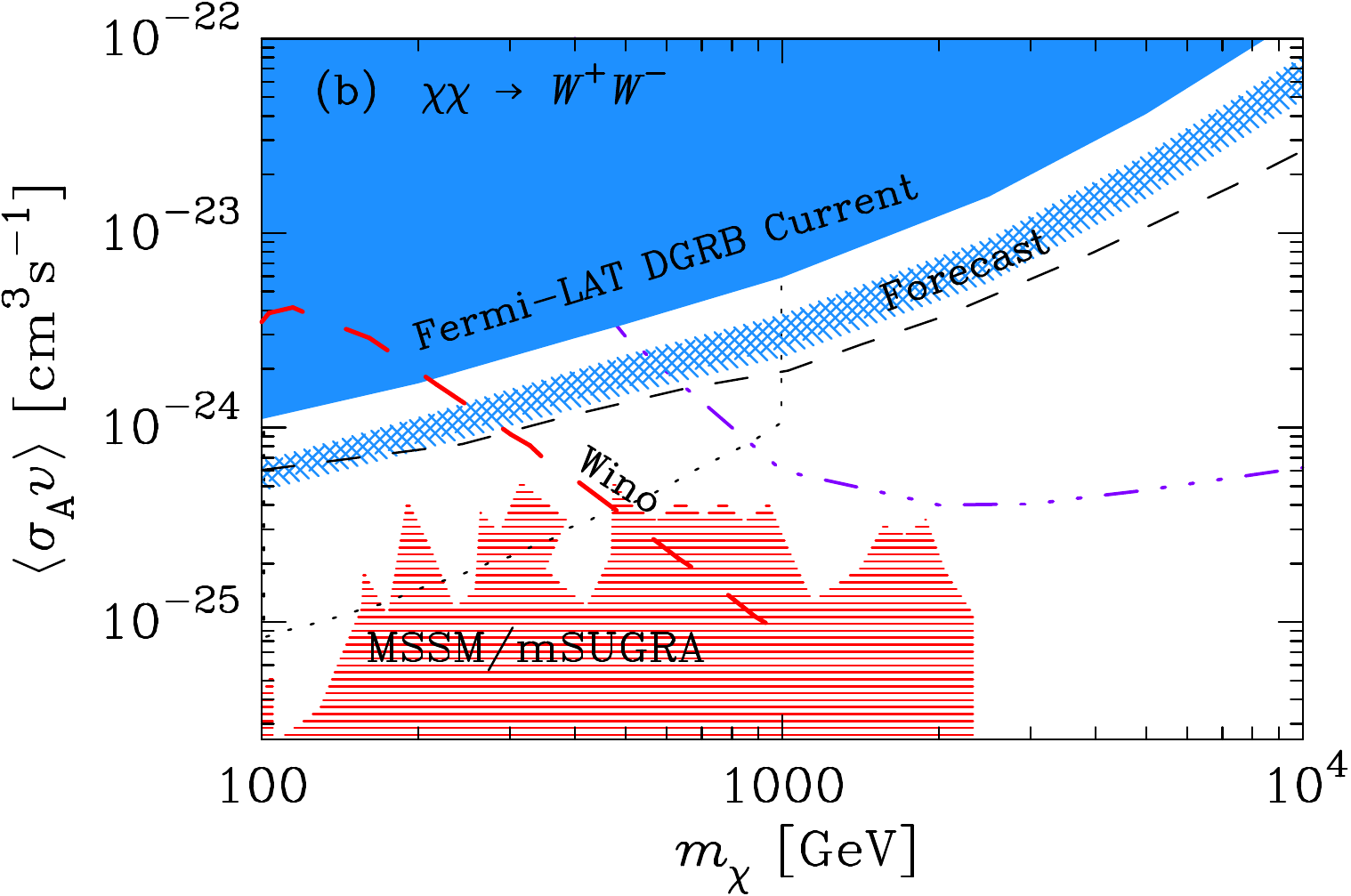} \\
\includegraphics[width=3.4truein]{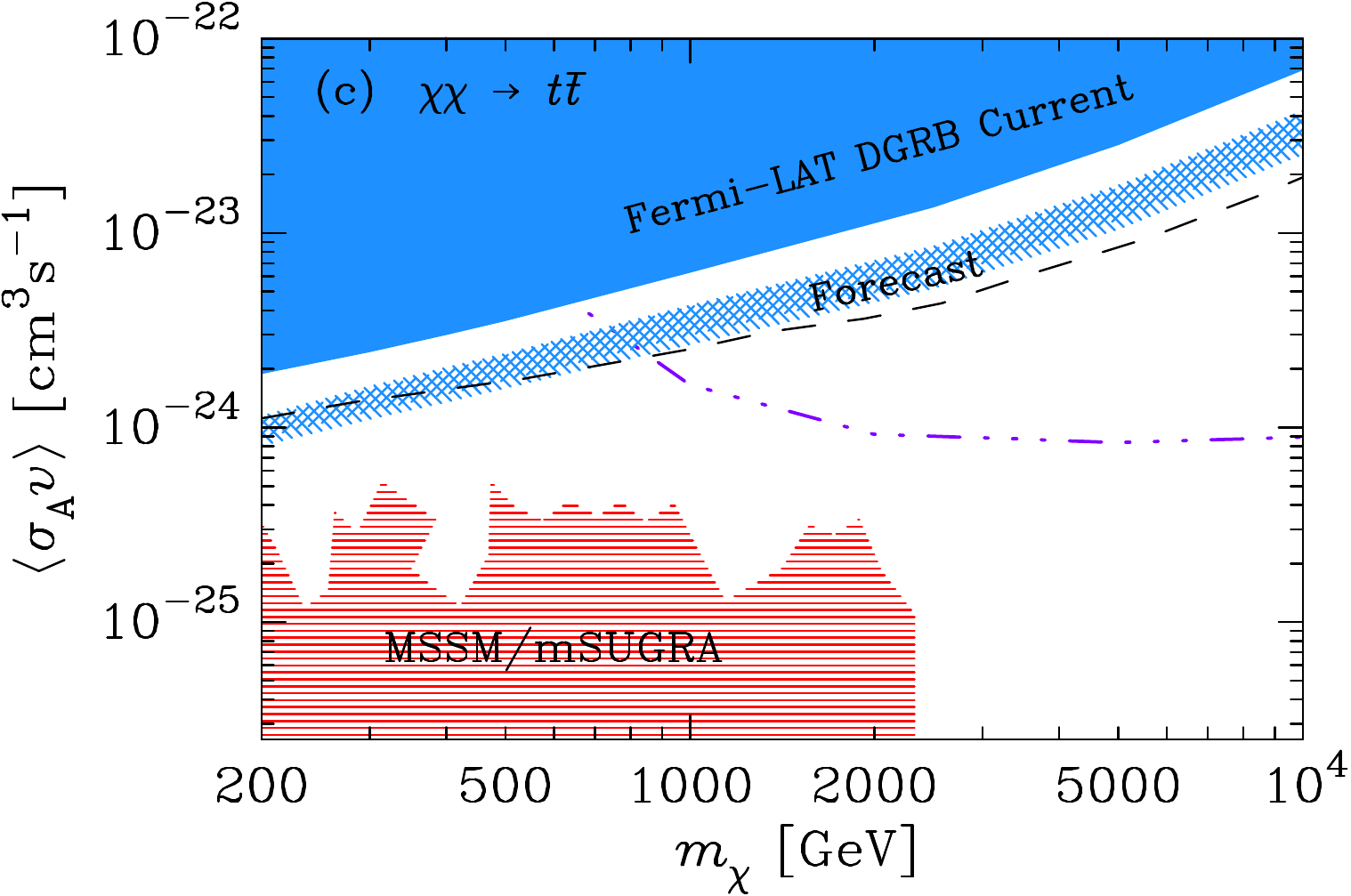} 
\hspace{0.3cm}
\includegraphics[width=3.4truein]{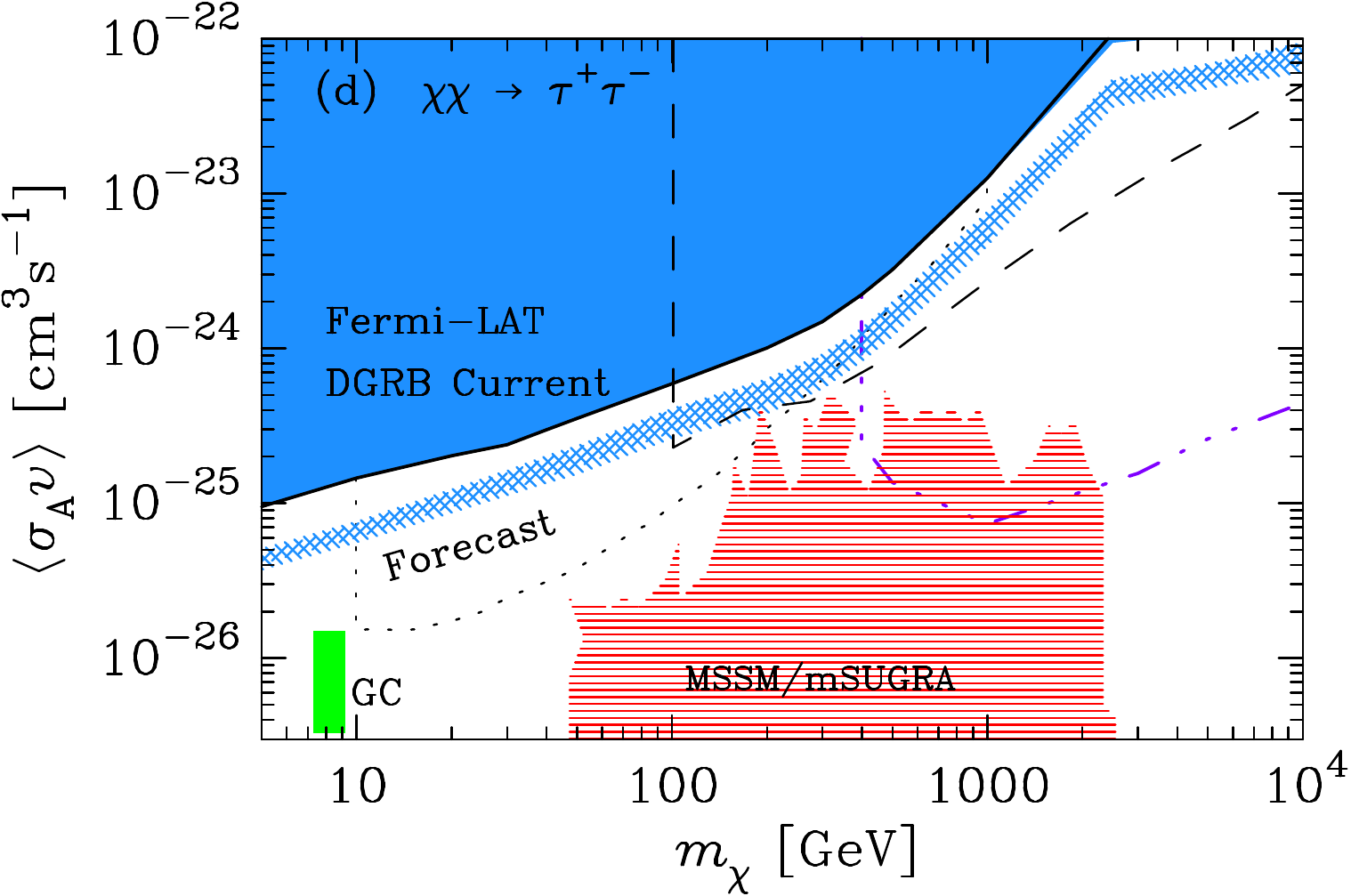}
\caption{Shown are our predictions for the Fermi-LAT sensitivity to
  constraints on dark matter in several canonical annihilation
  channels: (a) $\chi\chi\rightarrow b\bar b$; (b)
  $\chi\chi\rightarrow W^+W^-$; (c) $\chi\chi\rightarrow t\bar t$; (d)
  $\chi\chi\rightarrow \tau^+\tau^-$.  The blue double and single
  hashed regions are the 68\% and 95\% C.L.\ predictions for 5-year
  Fermi-LAT sensitivity, respectively. Also shown is the limit from
  the current Fermi-LAT DGRB spectrum (solid blue). In panel (a), we
  also show the current constraints from Fermi-LAT observations of
  Draco (thick dotted line) \cite{Abdo:2010ex}.  In panels (a), (b)
  and (d), we show constraints from the stacking of dwarf galaxies
  (thin dotted line) \cite{Ackermann:2011wa}.  In the $W^+W^-$
  channel, panel (b), the cross section versus mass for a nonthermal
  winolike neutralino is shown.  The winolike dark matter of the
  PAMELA signal at $m_\chi \sim 200\rm\ GeV$
  \cite{Grajek:2008jb,*Kane:2009if} is disfavored by the current
  constraints and will be further constrained in the forecast
  spectrum.  The `MSSM/mSUGRA' red-striped region is the expected
  cross sections for a sampling of points in supersymmetric parameter
  space~\cite{Bergstrom:2010gh}. The dashed line is a limit from the
  $3^\circ\times 3^\circ$ in the Galactic
  Center~\cite{Cirelli:2009dv}. In the $\tau^+\tau^-$ plot, panel (d),
  the green box is a region that could be consistent with an excess in
  the spectrum toward the Galactic Center~\cite{Hooper:2010mq}. In all
  panels, the triple-dot-dashed line is the limit from the HESS
  observation of the Galactic Center~\cite{Abramowski:2011hc} for the
  respective channels in the case of a Navarro-Frenk-White halo
  profile, from
  Ref.~\cite{Abazajian:2011ak}.  \label{annihilationplots}}
\end{figure*}

Figure~\ref{annihilationplots} demonstrates the forecast for four
canonical dark matter annihilation channels and how they compare to
expected minimal supersymmetric model (MSSM) and minimal supergravity
(mSUGRA) dark matter cross sections. The cross sections are from a
scan in MSSM and mSUGRA parameter space by
Ref.~\cite{Bergstrom:2010gh}. Note that our predicted constraints from
Fermi-LAT's observations of the DGRB are comparable to current
constraints Fermi-LAT observations of the Galactic Center
\cite{Cirelli:2009dv}, shown as a dashed line in
Fig.~\ref{annihilationplots}.  For comparison, in
Fig.~\ref{annihilationplots} we also show constraints from the
Fermi-LAT~ analysis of stacked dwarf galaxies
\cite{GeringerSameth:2011iw,Ackermann:2011wa}.  The constraints shown
are from Ref.~\cite{Ackermann:2011wa}.  In all relevant panels, we
also show constraints from the HESS observation of the Galactic
Center~\cite{Abramowski:2011hc} as calculated in
Ref.~\cite{Abazajian:2011ak}.

It has been proposed that a nonthermally-produced winolike dark
matter annihilation could lead to the positron excess signal in
PAMELA, which requires wino dark matter masses of $100{\rm\
  GeV}\lesssim m_\chi \lesssim 200\rm\ GeV$
\cite{Grajek:2008jb,*Kane:2009if}.  As shown in
Fig.~\ref{annihilationplots}(b), these models are disfavored by the
current constraints, and the forecast spectrum will be significantly
more sensitive to this model.

The standard thermal relic weakly-interacting dark matter annihilation
cross section is $\sim 3\times10^{-26}\rm\ cm^{3}\ s^{-1}$
\cite{Zeldovich65,*Steigman:1979kw,*Scherrer:1986}, which corresponds
roughly with the broader MSSM/mSUGRA region. The plot of dark matter
annihilating into $\tau^+\tau^-$, Fig.~\ref{annihilationplots}(d),
contains a region in parameter space which has recently been claimed
to be consistent with the morphology and spectrum of excess gamma-ray
flux towards the Galactic Center \cite{Hooper:2010mq}, though such a
signal is also consistent with emission from stellar
clusters~\cite{Abazajian:2010zy}. The width of the forecast region
follows from the increasing width of the DGRB prediction with
energy. Our model predicts a factor of 2 $-$ 3 (95\% C.L.) improvement
in the sensitivity of the DGRB measurement to dark matter.

\subsection{Diffuse Emission from Annihilation of Dark Matter into
  Leptonic Modes and IC}

\begin{figure*}[t]
 \includegraphics[width=3.4truein]{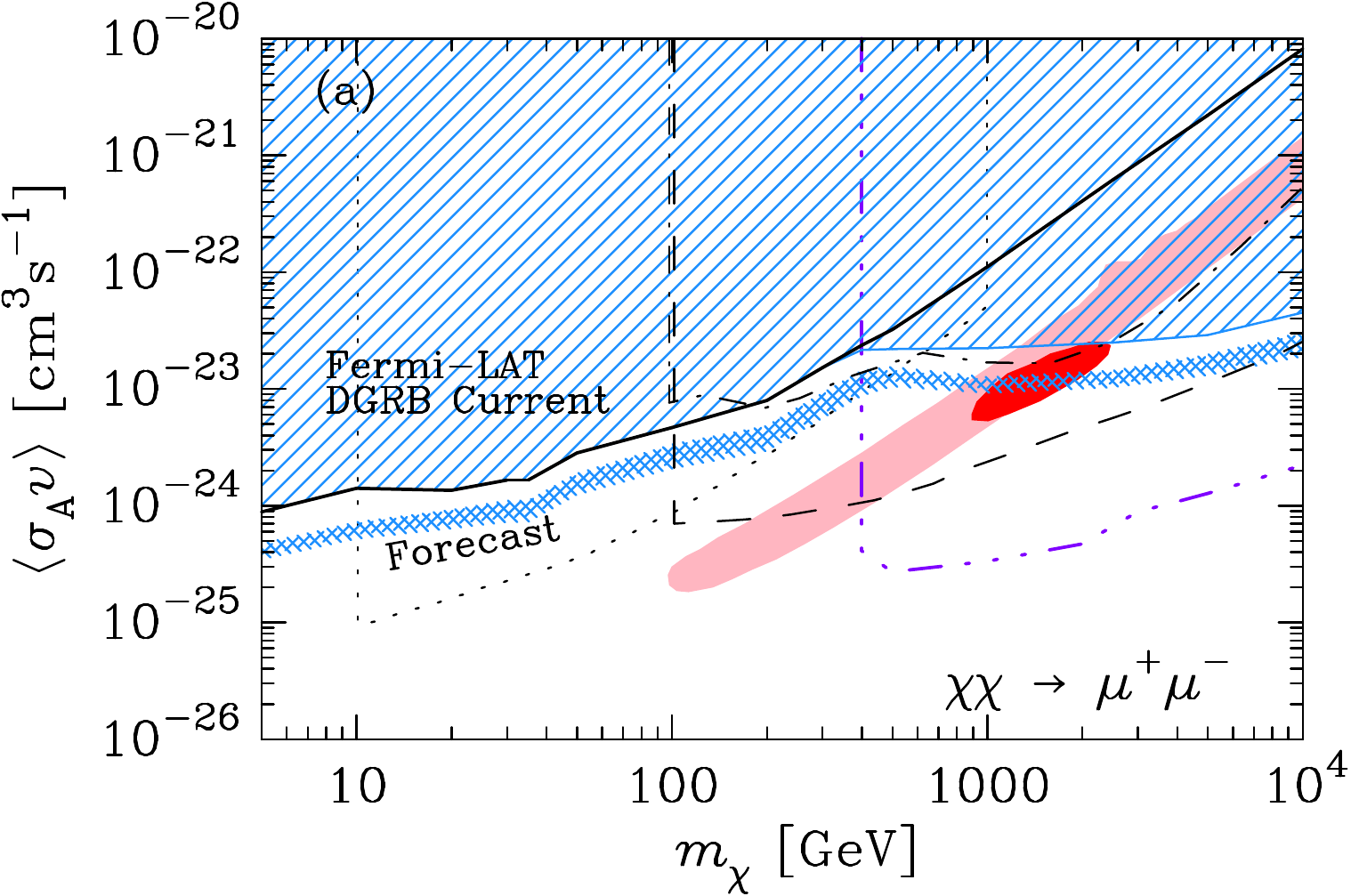}\hspace{0.3cm}
 \includegraphics[width=3.4truein]{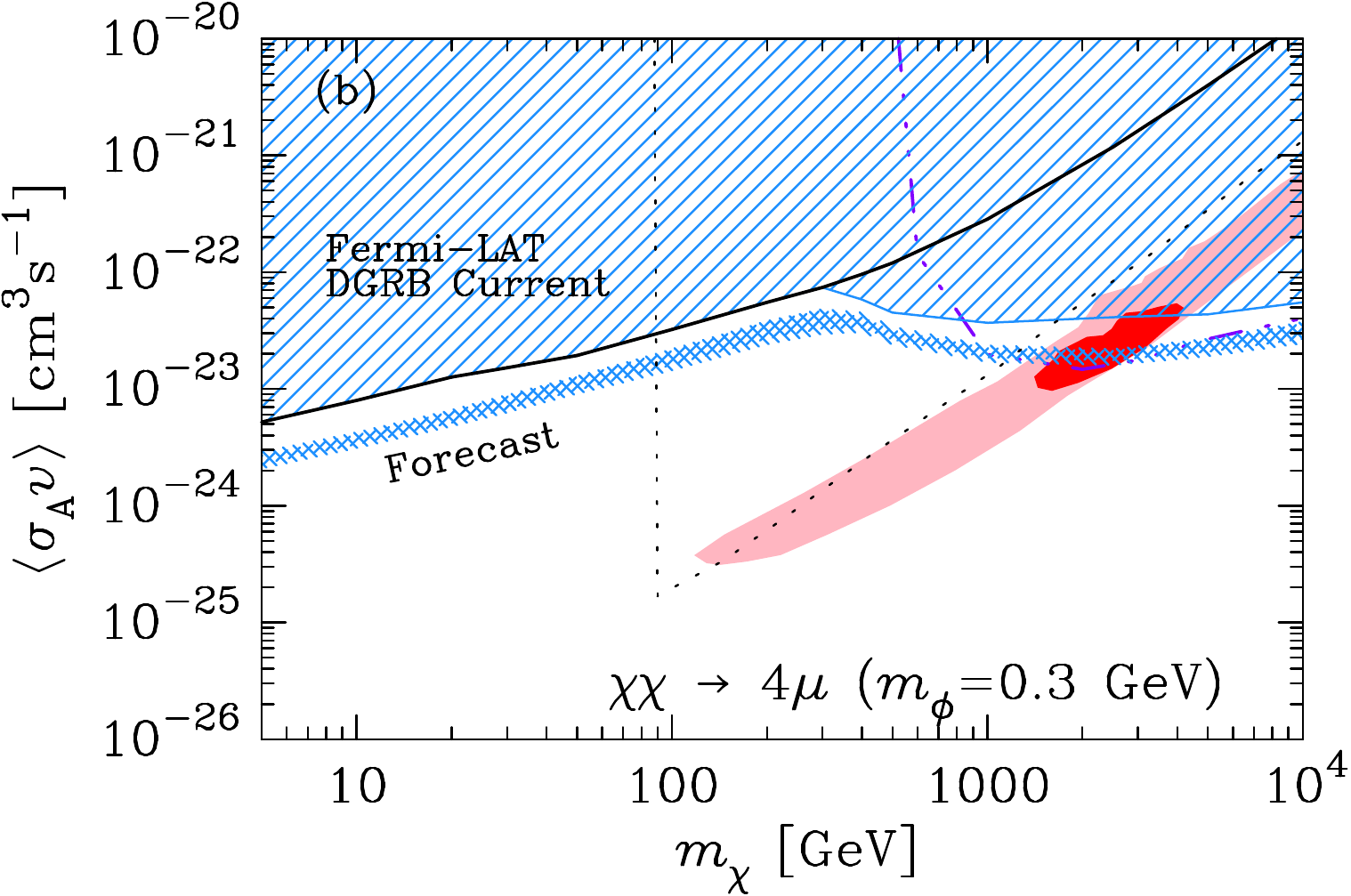}
 \caption{Shown in panel (a) are our predictions for Fermi-LAT
   sensitivity to $\mu^{+}\mu^{-}$ channel dark matter annihilation.
   The dot-dashed line is the 95\% C.L.\ limit on prompt and IC emission
   from Ursa Minor \cite{Abdo:2010ex}. In this panel, we show the
   constraint from the stacking of dwarf galaxies (thin dotted line)
   \cite{Ackermann:2011wa}.  In panel (b), dark matter annihilation
   into four muons via intermediate 0.3 GeV scalar particles
   $\phi$. The blue double and single hashed regions are the 68\% and
   95\% C.L.\ predictions for 5-year Fermi-LAT sensitivity,
   respectively. Also included is the limit from the current Fermi-LAT
   DGRB spectrum (striped blue region).  The solid black line shows
   where the exclusion would be without the IC contribution. The light
   pink shaded region is consistent with a dark matter interpretation
   of the PAMELA signal and the dark red shaded region is consistent
   with a dark matter interpretation of the Fermi-LAT $e^{+}/e^{-}$
   feature~\cite{Meade:2009rb,Papucci:2009gd}.  The triple-dot-dashed
   line is the limit from the HESS observation of the Galactic
   Center~\cite{Abramowski:2011hc} for the respective channels in the
   case of a Navarro-Frenk-White halo profile, from
   Ref.~\cite{Abazajian:2011ak}.  In panel (b), the dotted line is a
   limit on dark matter annihilation from radio synchrotron from the
   Galactic Center~\cite{Meade:2009rb}. The dashed line is the 99\% C.L.
   limit from the $3^\circ\times 3^\circ$ region toward the Galactic
   Center~\cite{Cirelli:2009dv}. \label{pamela}}
\end{figure*}

Dark matter that annihilates into leptons has been proposed as an
explanation for an excess in cosmic-ray positrons seen by the PAMELA
Collaboration~\cite{Adriani:2008zr} and a feature in the cosmic-ray
$e^+/e^-$ spectrum from the
Fermi-LAT~\cite{Abdo:2009zk,*Grasso:2009ma,*Slatyer:2009yq}. The DGRB
is forecast to be sensitive to direct $\mu^+\mu^-$ production models
that fit the PAMELA excess and Fermi-LAT $e^+/e^-$ feature, such as
Asymmetric Dark Matter~\cite{Falkowski:2011xh,*Cai:2009ia}.  Such
direct annihilation models are already strongly disfavored by
constraints of observations by the Galactic Ridge by
HESS~\cite{Aharonian:2006au,Bertone:2008xr,Abazajian:2010sq},
observations by Fermi-LAT toward the Galactic
Center~\cite{Cirelli:2009dv}, and HESS toward the Galactic
Center~\cite{Abramowski:2011hc,Abazajian:2011ak}.  The latter two
constraints shown in Fig.~\ref{pamela}.

It has also been suggested that Sommerfeld-enhanced dark matter
annihilation into four leptons, via a light mediator particle, could
explain these signals. Therefore, we also consider how an improved
DGRB will affect these leptonic dark matter annihilation channels.
Note that such models are also constrained by detailed calculations of
the relic abundance from dark matter production in the early Universe
and halo shapes \cite{Feng:2009hw}, as well as distortions of the CMB
spectrum \cite{Galli:2009zc,*Zavala:2009mi}.  Recent work finds that
several reasonable parameter choices in models of Sommerfeld-enhanced
dark matter for PAMELA and the Fermi-LAT $e^+/e^-$ spectral feature
avoid these limits~\cite{Finkbeiner:2010sm}.  These
Sommerfeld-enhanced models are partially constrained by the current
limits and will be further constrained by our forecast limits, as
shown in Fig.~\ref{pamela}(b).

Dark matter annihilation into leptons produces fewer photons than the
quark or gauge boson channels. However, such annihilations do produce
highly-boosted electrons which undergo IC scattering on the CMB and
starlight which is then observable in gamma-rays. We calculate this IC
contribution as in Ref.~\cite{Cirelli:2009dv}. To be conservative, we only
include the scattering from the CMB, not from starlight or the IR
background.

The spectrum of IC photons coming from one dark matter annihilation is
given by~\cite{Cirelli:2009dv,Profumo:2009uf}
\begin{equation}
\frac{dN}{dE}=\frac{1}{E}\int_{m_{e}}^{m_{\chi}}d\epsilon\frac{\mathcal{P}(E,\epsilon)}
{\dot{\mathcal{E}}(\epsilon)}Y(\epsilon), 
\end{equation}
where $\mathcal{P}(E,\epsilon)$ is the differential power emitted into
photons of energy $E$ by an electron with energy $\epsilon$,
$\dot{\mathcal{E}}$ denotes the total rate of electron energy loss due
to IC scattering, and $Y(\epsilon)$ is the number of electrons
generated with energy larger than $\epsilon$ in one annihilation. To
get the energy of the annihilation products, we use the software
{\sc PYTHIA}~\cite{Sjostrand:2006za}. In the Thomson limit, which is a very good
approximation for CMB photons, one obtains
\begin{eqnarray}
  \dot{\mathcal{E}}(\epsilon)&=&\frac{4}{3}\sigma_{T}\gamma^{2}\int_{0}^{\infty}d\epsilon'\epsilon' n(\epsilon')\\
  \mathcal{P}(E,\epsilon)&=&\frac{3\sigma_{T}}{4\gamma^{2}}E\int_{0}^{1}dy\frac{n(\epsilon'(y))}{y}\nonumber\\
  &&\times\left(2y\ln y+y+1-2y^2\right),
\end{eqnarray}
where $\gamma$ is the Lorentz factor of the electron, $\epsilon'$ is
the energy of the initial CMB photon, $y\equiv
E/(4\gamma^{2}\epsilon')$, $\sigma_{T}\simeq 0.665\rm\ barn$ is the
Thomson cross section, and the radiation density of CMB photons at
$T_{\rm CMB}\simeq 2.725\rm\ K$ is given by
\begin{equation}
n_{\rm
  CMB}(\epsilon')=\frac{\epsilon'^{2}}{\pi^{2}}\frac{1}{\exp(\epsilon'/T_{\rm
    CMB})-1}.
\end{equation}
The flux coming from the IC contribution can then be calculated as in
Eqs.~(\ref{DMannfluxeqn})-(\ref{fando}). 

It is important to note that we are neglecting the diffusion of the
electrons and positrons from the dark matter annihilation to the point
where IC scattering occurs. As stated in Ref.~\cite{Cirelli:2009dv},
this is a good approximation in regions away from the Galactic Center,
as we consider in this work. For an analysis where diffusion is
included, see Ref.~\cite{Papucci:2009gd}.  To be more explicit, in
Ref.~\cite{Pieri:2009je}, it was shown that the electron/positron flux
produced by dark matter annihilations was only slightly modified by
diffusion effects for distances from the Galactic Center larger than 8
kpc. Corrections were found to reach a factor of 2 only at the lowest
energies around 100 MeV (see their Fig. 14).  Therefore, diffusion
effects are not significant in our calculation of the IC component.
Note that the IC component is modeled in the measurement if the DGRB
independent of the observation~\cite{Abdo:2010nz}, so there is no
accidental subtraction of a potential IC signal.

Our calculations for the standard $\mu^+\mu^-$ leptonic channel are
shown in Fig.~\ref{pamela}(a). Also shown are the regions which are
consistent with the PAMELA excess and Fermi-LAT $e^+/e^-$ spectral
feature from Ref.~\cite{Meade:2009rb}, modified for a higher local
dark matter density of our minimal Einasto halo model ($\rho_\odot =
0.385\rm\ GeV\ cm^{-3}$), and the local boost of $B(r=R_\odot) = 1.57$
[Eq.~\eqref{boostatr}] from substructure.  For this boost, we use the
same parameters for the subhalo PDF as our subhalo boost annihilation
signal. Such models are already highly constrained by several
gamma-ray observations as shown in Fig.~\ref{pamela}(a), and such
models will be further constrained with our forecast DGRB
sensitivity. Note that the IC contribution improves the bounds by
several orders of magnitude. The IC gamma-ray flux contribution peaks
at much lower energies than the prompt component for a given dark
matter particle mass, which is why the width of the forecast region
decreases when the IC component becomes important. Also shown in
Fig.~\ref{pamela} are complementary limits on the cross section of
$\mu^+\mu^-$-channel annihilating dark matter from other work.  Shown
in Fig.~\ref{pamela}(b) is dark matter annihilation into four muons
via intermediate scalars $\phi$ with 0.3 GeV masses. Even in this
less-constrained case, our forecast is that the Fermi-LAT measurement
of the DGRB will have the sensitivity to detect or rule out a portion
of the parameter space the dark matter interpretations of PAMELA.

\subsection{Diffuse Emission from Decaying Dark Matter}

\begin{figure}[t]
\begin{center}
\includegraphics[width=3.4truein]{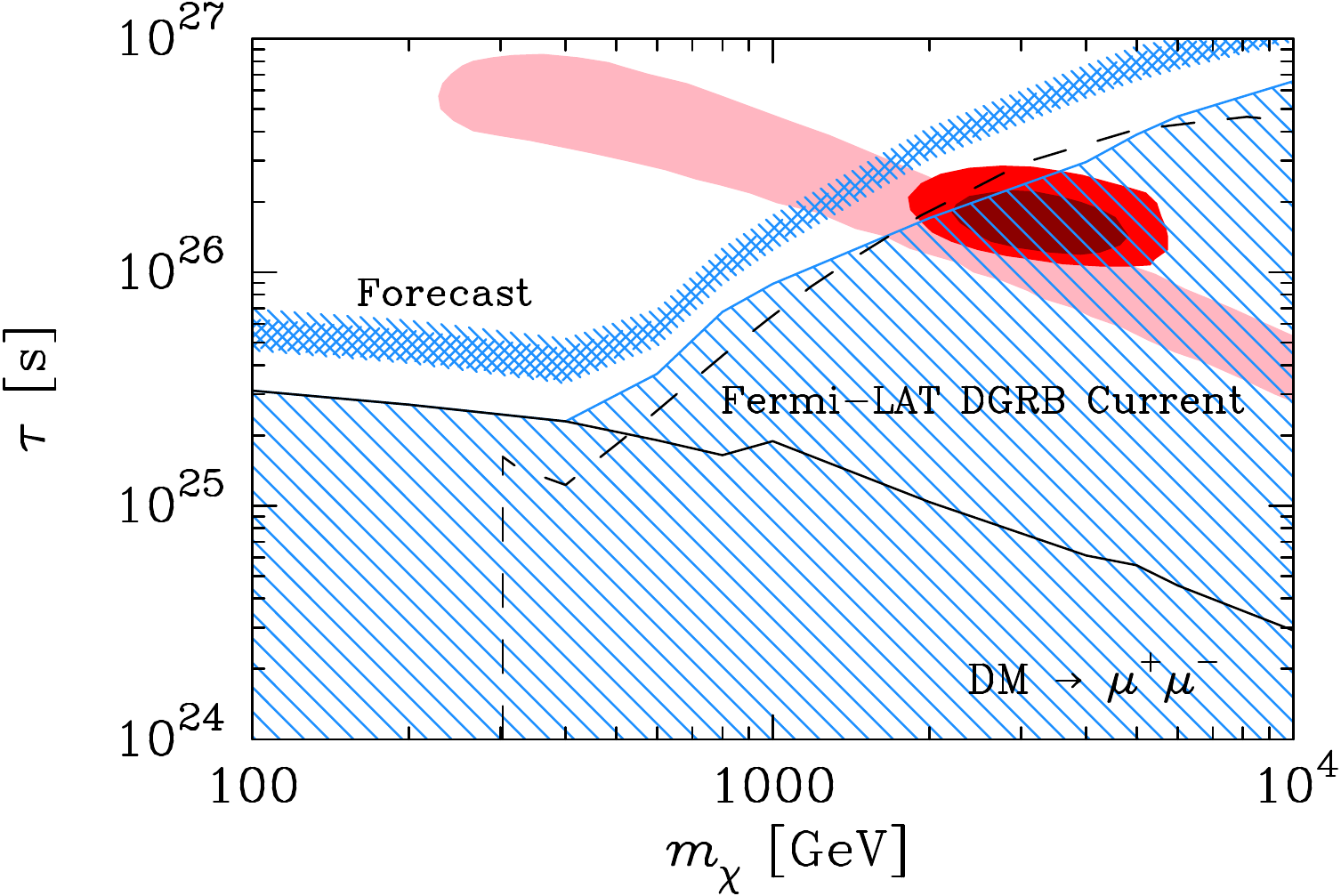}
\end{center}
\caption{Our predictions for Fermi-LAT sensitivity to dark matter
  decaying into $\mu^{+}\mu^{-}$. The dark and light blue hashed
  regions are the 68\% and 95\% C.L.\ predictions for 5-year Fermi-LAT
  sensitivity, respectively. Also included is the limit from the
  current Fermi-LAT DGRB spectrum (blue striped region). The solid
  black line shows where the exclusion would be without the IC
  contribution. The light pink shaded region is consistent with a dark
  matter interpretation of the PAMELA signal, and the dark red shaded
  regions are consistent with a dark matter interpretation of the
  Fermi-LAT $e^{+}/e^{-}$ feature at 3- and
  5-$\sigma$~\cite{Papucci:2009gd}. The dashed line is from
  constraints on prompt and IC emission from dark matter annihilation
  made from Fermi-LAT observations of the Fornax cluster of
  galaxies~\cite{Dugger:2010ys}.\label{decayplot}}
\end{figure}

Decaying dark matter is another, less constrained, possibility for the
source of the positron fraction signal seen in PAMELA and the
Fermi-LAT $e^+/e^-$ spectral feature~\cite{Cirelli:2009dv}.  To
calculate the flux from decaying dark matter with lifetime $\tau$ and
photon spectrum $dN_\gamma/dE$, the procedure is very similar to the
annihilating dark matter case. However, in Eq.~(\ref{rhosqinteqn}),
$\rho^2\rightarrow\rho$ and there is no boost factor due to the lack
of decay enhancement with density. Here, Eq.~(\ref{DMannfluxeqn})
becomes
\begin{equation}
  \frac{d\Phi_\gamma}{dEd\Omega}=\frac{1}{\tau}\frac{\mathcal{J}_{\rm
      AGC}}{\rm J_0} \frac{1}{4\pi m_\chi} \frac{dN_\gamma}{dE}\label{DMdecfluxeqn}.
\end{equation}
Similarly, to calculate the extragalactic flux from decaying dark
matter, in Eq.~(\ref{EGdecay}) replace
\begin{equation}
  \frac{\langle\sigma_A v\rangle}{2}\frac{(f_{\rm DM} \Omega_m)^2\rho_{\rm crit}^2}{m_\chi^2}\rightarrow \frac{1}{\tau}\frac{(f_{\rm DM} \Omega_m)\rho_{\rm crit}}{m_\chi}
\end{equation}
and drop all boost factors.

Fig.~\ref{decayplot} shows how the predicted DGRB value will constrain
the lifetime of a dark matter particle which decays into
$\mu^+\mu^-$. The improved constraint will have the sensitivity to
exclude or detect the decaying dark matter interpretation of the
Fermi-LAT $e^+/e^-$ feature and should provide strong limits on an
interpretation of the PAMELA
excess~\cite{Adriani:2008zr,Papucci:2009gd}.  The DGRB limit is
comparable to and is forecast to be more sensitive than the limits on
decaying dark matter from Fermi-LAT observations of clusters of
galaxies~\cite{Dugger:2010ys}, as shown in Fig.~\ref{decayplot}.

\subsection{Comparison to Direct Dark Matter Detection Limits on Light
Dark Matter}

\begin{figure}[t]
\begin{center}
\includegraphics[width=3.4truein]{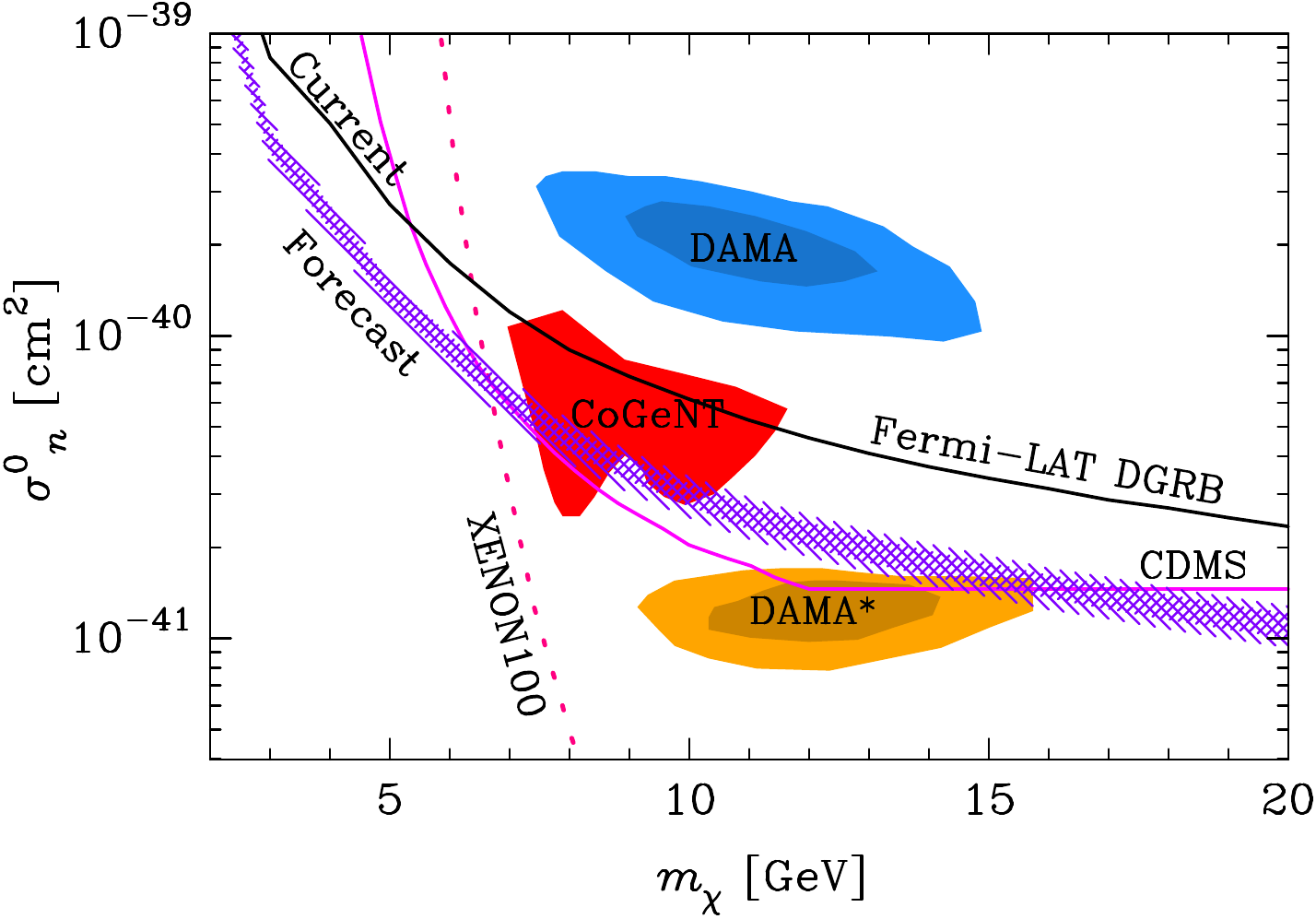}
\end{center}
\caption{Our predictions for Fermi-LAT sensitivity to light dark
  matter coupled via Higgs-like
  couplings~\cite{Andreas:2008xy,*Arina:2010rb}. The dark and light
  purple hashed regions are the 68\% and 95\% C.L.\ predictions for
  5-year Fermi-LAT sensitivity, respectively. Also included is the
  limit from the current Fermi-LAT DGRB spectrum (black line). To the
  right of the dotted pink line is exclusion from the XENON100
  Collaboration~\cite{Aprile:2011hx}. The red ``CoGeNT'' region is
  consistent with the findings of the CoGeNT
  Collaboration~\cite{Aalseth:2010vx}. The blue upper ``DAMA'' region is the
  DAMA signal without channeling~\cite{Savage:2010tg}. The orange
  lower ``DAMA*'' region is the DAMA signal if channeling is
  included~\cite{Kopp:2009qt,Feldstein:2009np}. The region labeled
  ``CDMS'' is excluded by that experiment's light dark matter search at
  95\% C.L.~\cite{Ahmed:2010wy}.  \label{lightDMplot}}
\end{figure}

Dark matter may be detected through two distinct methods:
indirect-detection experiments seek the annihilation or decay products
from dark matter in the Universe, while direct-detection experiments
look for the recoil of heavy nuclei after their collision with a dark
matter particle from our Galactic halo. In general, the interaction
cross section between annihilating dark matter is not simply related
to the interaction cross section between dark matter and nucleons. A
few recent direct-detection experiments, however, have seen signals
that could be caused by a light dark matter particle, including
DAMA~\cite{Bernabei:2008yi,Feldstein:2009np,Kopp:2009qt,Savage:2010tg},
CoGeNT~\cite{Aalseth:2010vx} and CDMS~\cite{Ahmed:2010wy}.

These light dark matter signals could be due to dark matter that
interacts through the exchange of Higgs
bosons~\cite{Burgess:2000yq,Andreas:2008xy,*Arina:2010rb}. For such a
dark matter candidate, the indirect annihilation cross section and
direct nuclear cross section are related by
\begin{eqnarray}
\!  \sigma_{\rm ind}(S S \rightarrow \bar{f} f) \frac{v}{c}&=&n_c\, \frac{\lambda_L^2}{\pi}\frac{m_f^2(m_S^2-m_f^2)^{3/2}}{m_h^4 m_S^3}\label{sigmaSannih}\\
\!  \sigma_{\rm dir}(S {N} \rightarrow S {N})&=&\frac{\lambda_L^2}{\pi}  \frac{\mu_r^2}{m^4_h m_{S}^2} f^2 m_{N}^2 \label{sigmaSdirectdet}\\
\!  \sum\frac{\sigma_{\rm ind}}{\sigma_{\rm dir}}\frac{v}{c}&=&\sum\frac{n_c m^2_f}{f^2 m^2_N}\frac{(m_S^2-m_f^2)^{3/2}}{\mu^2_r m_S},
\label{ratioS}
\end{eqnarray}
where $m_h$ is the Higgs mass, $\lambda_L$ is the dark matter-Higgs
coupling, $n_c =3(1)$ for quarks (leptons), $\mu_r$ is the dark
matter-nucleon reduced mass, $f\sim 0.3$, and $m_S$ is the dark matter
mass~\cite{Andreas:2008xy,*Arina:2010rb}. The sum in
Eq.~(\ref{ratioS}) is over all annihilation products, which are
dominated by the $b$-quark, the $c$-quark, and the
$\tau$-lepton. Through this ratio, we can relate our predicted DGRB
limits on indirect detection into limits on direct-detection
experiments. For such a Higgs-mediated dark matter model, we compare
our projected limit to the findings of several direct-detection
experiments in Fig.~\ref{lightDMplot}.

The limit on the dark matter-nucleon cross section as found by our
DGRB forecast is competitive with the limits by the XENON100
Collaboration in the lowest mass range~\cite{Aprile:2011hx}. The current
Fermi-LAT DGRB values rule out the DAMA region without channeling and
some of the region consistent with a dark matter interpretation of
CoGeNT. After a Fermi-LAT 5-year run, we forecast the DGRB spectrum to
have the sensitivity to exclude most of the CoGeNT region consistent
with dark matter interpretations in the spin-independent case. This is
complementary to the findings of direct-detection experiments since
the DGRB indirect-detection limits tend to exclude lower dark matter
masses than direct-detection experiments.

\section{Conclusions}
The DGRB as measured by Fermi-LAT is one of the most powerful
constraints on annihilating weak-scale particle dark matter.  We show
that the likely resolution of blazars into point sources by
Fermi-LAT---and their automatic removal from the DGRB
measurement---will enhance the sensitivity of the DGRB to dark matter
annihilation by a factor of 2 to 3 (95\% C.L.), depending on the
channel, mass scale, and true realization of the blazar distribution
and SED sequence.

We find the forecast dark matter sensitivity of the DGRB observation
to both prompt and inverse-Compton photon emission to be comparable in
sensitivity with other limits on annihilating weak-scale dark
matter.  The DGRB is forecast to be comparable to current limits from
Fermi-LAT observations toward the Galactic
Center~\cite{Cirelli:2009dv}, individual dwarf
galaxies~\cite{Abdo:2010ex}, and, in the case of decaying dark matter,
observations of clusters of galaxies~\cite{Dugger:2010ys}.  This
sensitivity makes the DGRB among the best methods of detecting or
constraining dark matter with the Fermi-LAT mission.  The forecast for
the DGRB projects it to be less constraining, for certain channels
and particle masses, than current stacked dwarf analyses with
Fermi-LAT~\cite{GeringerSameth:2011iw,Ackermann:2011wa} and
observations of the Galactic Center by
HESS~\cite{Abramowski:2011hc,Abazajian:2011ak}.

The future resolution and reduction of the DGRB into blazar point
sources highlights and enhances the possibility that the Fermi-LAT
experiment will either detect or constrain the dark matter in a robust
yet conservative way.

\begin{acknowledgments}
  We would like to thank P.  Agrawal, J. Beacom, Z.\ Chacko,
  J.\ McEnery and N.\ Weiner for useful discussions.  KNA and JPH are
  supported by NSF Grant No.\ 07-57966 and NSF CAREER Award
  No.\ 09-55415. This work has been partially supported by MICNN,
  Spain, under contracts FPA 2007-60252 and Consolider-Ingenio CPAN
  CSD2007-00042 and by the the Comunidad de Madrid through Proyecto
  HEPHACOS ESP-1473.  S.B. acknowledges support from the CSIC under
  Grant No.\ JAE-DOC.
\end{acknowledgments}

\bibliography{master}

\end{document}